\documentstyle[12pt,subeqn,subeqnarray,fullpage]{article}

\def\lsim{\:\raisebox{-0.5ex}{$\stackrel{\textstyle<}{\sim}$}\:}
 
\def\be{\begin{equation}}
\def\ee{\end{equation}}
\def\beq{\begin{eqnarray}}
\def\eeq{\end{eqnarray}}
\def\bsubeq{\begin{subeqnarray}}
\def\esubeq{\end{subeqnarray}}
\def\s{\subequations}
\def\es{\endsubequations}
\def\R{${R\!\!\!/}_p$}
\def\E{${E\!\!\!\!/}_T$}

\def\L{${L\!\!\!/}$}

\begin{document}
\begin{flushright}
TIFR/TH/97-60
\end{flushright}
\bigskip\bigskip
\begin{center}
\Large{\bf Supersymmetric theories with explicit R-parity
violation}\footnote{Plenary talk at the Pacific Particle Physics
Phenomenology Workshop, Seoul, 31 Oct. -- 2 Nov., 1997} \\[1cm]
\large{Probir Roy} \\
Tata Institute of Fundamental Research, Mumbai, India
\end{center}
\bigskip\bigskip
\begin{enumerate}
\item[{}] $\bullet$ {\bf $R_p$ and motivation for explicit \R}
$\bullet$ {\bf MSSM $\rightarrow$ \R \ MSSM} \\ 
$\bullet$ {\bf Bounds on \R \ couplings} $\bullet$ {\bf
Collider-specific phenomenology} \\ $\bullet$ {\bf Cosmological
implications} $\bullet$ {\bf Summary and outlook} $\bullet$ {\bf References}
\end{enumerate}
\bigskip\bigskip

\noindent $\bullet$ {\bf $R_p$ and motivation for explicit \R}
\smallskip

The superfield spectrum of the Minimal Supersymmetric Standard Model
$MSSM$ [1] can be written in usual notation as 
\[
Q_i = \left(\matrix{U_i \cr D_i}\right), \ L_i = \left(\matrix{N_i \cr
E_i}\right), \ H_u = \left(\matrix{H^+_u \cr H^0_u}\right), \ H_d =
\left(\matrix{H^0_d \cr H^-_d}\right), \ \bar U_i, \bar D_i, \bar E_i 
\]
and its superpotential as 
\be
W_{MSSM} = h^u_{ij} Q_i \cdot H_i U_j + h^d_{ij} Q_i \cdot H_d \bar
D_j + h^e_{ij} L_i \cdot H_d \bar E_j + \mu H_u \cdot H_d.
\ee
(1) has been constructed so as to conserve $R$-parity (even for particles,
odd for sparticles)
\[
R_p \equiv (-1)^{3B+L+2s}, 
\]
$B,L$ and $s$ being baryon no., lepton no. and spin.  This is
equivalent to the conservation of matter parity which transforms the
superfields as 
\beq
(L_i, \bar E_i, Q_i, \bar U_i, \bar D_i) &\rightarrow& -(L_i, \bar
E_i, Q_i, \bar U_i, \bar D_i), \nonumber \\[2mm]
(H_u, H_d, V_\gamma, V_Z, V^+_W, V^-_W) &\rightarrow& (H_u, H_d, V_\gamma, V_Z,
V^+_W, V^-_W). \nonumber
\eeq

Three spectacular consequences ensue from $R_p$-conservation.  First,
each vertex involving sparticles has them in a pair.  Second, the
lightest supersymmetric particle (LSP $\tilde\chi^0_1$: 
a weakly interacting neutralino, escaping through detectors, being
neutral on
cosmological grounds) is an excellent candidate for cold dark
matter.  Finally, each sparticle, pair-produced in a collider,
would decay within the detector into particles and $\tilde\chi^0_1$,
the latter being characterized by a hard (\E) signature.

Recall that $R_p$-conservation in supersymmetric theories was
phenomenologically postulated [2] by the need to avoid catastrophic
proton decays induced by \R \ vertices.  However, such decays need to
violate both $B$ and $L$.  So, in place of $R_p$, it suffices to have
another discrete symmetry such as $B$-parity transforming the
superfields as 
\[
(Q_i,\bar U_i,\bar D_i) \rightarrow -(Q_i,\bar U_i, \bar D_i),
\]
\[
(L_i,\bar E_i, H_u, H_d, V_\gamma,V_Z,V^+_W,V^-_W) \rightarrow (L_i,\bar
E_i, H_u, H_d, V_\gamma,V_Z,V^+_W,V^-_W).
\]
Analogously, one can have an $L$-parity instead.  Since our presently
scanty knowledge of high scale theories is unable [3] to discriminate
between various possible residual discrete symmetries at low energies,
it is quite reasonable to consider [4] \R \ (but $L$- or $B$-conserving)
interactions near weak-scale energies.  In fact, supersymmetric grand
unified theories have been constructed [5] which have \R \ interactions
present at low energies.  There are even stringy scenarios [6] which
predict them.  On the other hand, a spontaneous violation of $R_p$
requires [7] an extra $SM$-singlet and hence a nonminimal spectrum; so
we will restrict ourselves to explicit $R_p$-violation.  The LSP, in
this scenario, is neither stable nor necessarily neutral.
\bigskip

\noindent $\bullet$ {\bf MSSM $\rightarrow$ \R \ MSSM}
\medskip

\nobreak
All renormalizable supersymmetric\footnote{Supersymmetry-breaking \R \
terms can occur in the Lagrangian in the form of trilinear scalar
products.  These have less spectacular effects than the supersymmetric
\R \ interactions to which we restrict ourselves.} \R \ interactions
are incorporated by the addition of new terms in the superpotential
\s
\be
W_{MSSM} \rightarrow W_{MSSM} + W_{{R\!\!\!/}_p},
\ee
\be
W_{{R\!\!\!/}_p} = {1\over2} \lambda_{[ij]k} L_i \cdot L_j \bar E_k +
\lambda'_{ijk} L_i \cdot Q_j \bar D_k + {1\over2}
\lambda^{\prime\prime}_{i[jk]} \bar U_i \bar D_j \bar D_k + \epsilon_i
L_i \cdot H_u,
\ee
\es
$i,j,k$ being flavor indices and square brackets on them testifying to
antisymmetry in the bracketted indices.   The last RHS term in (2) can
be ``absorbed'' in the $\mu$-term of (1) by the transformation $H_d
\rightarrow H_d' = (\mu^2 + \epsilon_j \epsilon_j)^{-1/2} (\mu H_d -
\epsilon_i L_i)$.  This will, in general, lead to \ \L \ terms in the
scalar potential and, unless one is careful, can lead to nonzero
sneutrino VEVs $\langle \tilde\nu_i\rangle \neq 0$.  Moreover, the
abovementioned ``absorption'' works at a certain energy scale.  At a
different energy the $\mu$ and $\epsilon_i$ parameters will be
different, on account of separate renormalization group evolutions,
leading to a ``reappearance'' of the $\epsilon_i L_i \cdot H_u$ terms in
$W$.  However, the effects of such terms are smaller [8] than the
bounds on the $\lambda$-, $\lambda'$-,
$\lambda^{\prime\prime}$-couplings that we discuss below and so
henceforth we will ignore the $\epsilon_i L_i \cdot H_u$ piece in (2b).

We still have 9 $\lambda$-type, 27 $\lambda'$-type and 9
$\lambda^{\prime\prime}$-type, i.e. 45 possible, independent extra, couplings.
The \R \ Lagrangian density reads 
\newpage
\beq
{\cal L}_{{R\!\!\!/}_p} &=& 
\lambda_{[ij]k} \Big[\tilde\nu_{iL} \bar
e_{kR} e_{jL} + \tilde e_{jL} \bar e_{kR} \nu_{iL} + \tilde
e^\star_{kR} \overline{(\nu_{iL})^C} e_{jL} - \tilde\nu_{jL} \bar
e_{kR} e_{iL} \nonumber \\[2mm] && - \tilde e_{iL}
\bar e_{kR} \nu_{jL} + \tilde e^\star_{kR} \overline{(\nu_{jL})^C}
e_{iL}\Big] + \lambda'_{ijk} \Big[\tilde\nu_{iL} \bar d_{kR}
d_{jL} + \tilde d_{jL} \bar d_{kR} \nu_{iL} \nonumber \\[2mm]
&& + \tilde d^\star_{kR} \overline{(\nu_{iL})^C} d_{jL} - \tilde e_{iL} \bar
d_{kR} u_{jL} - \tilde u_{jL} \bar d_{kR} e_{jL} - \tilde
d^\star_{kR} \overline{(e_{iL})^C} u_{jL}\Big] 
+ \lambda^{\prime\prime}_{i[jk]} \epsilon_{\alpha\beta\gamma} 
\nonumber \\[2mm] 
&& \Big[\tilde u^\star_{iR\alpha} \bar
d_{kR\beta} d^C_{jR\gamma} + \tilde d_{jR\beta} \bar
e_{kR\gamma} u^C_{iR\alpha} + \tilde d^\star_{kR\gamma}
\overline{(u_{iR\alpha})^C} d_{jR\beta}\Big] + h.c.
\eeq
Generally, except in a specific case, we take the couplings to be
real.  Moreover, in analogy with the Yukawa couplings of MSSM, we
anticipate a generation hierarchy in their strengths with the third
generation ones expected to be the largest.  Furthermore, the simultaneous
presence of $\lambda$ or $\lambda'$ {\it and} $\lambda^{\prime\prime}$
type of couplings is extremely unlikely given the lack of observation
of either baryon nonconserving nucleon decay or double-nucleon
annihilation in nuclei.  Indeed, these constraints imply [8]
\s
\be
|\lambda'_{11k} \lambda^{\prime\prime}_{11k}| \lsim 2 \times 10^{-27}
\left({m_{\tilde k} \over 100 \ {\rm GeV}}\right)^2, 
\ee
\be
|\lambda_{233} \lambda^{\prime\prime}_{112}| \lsim 10^{-14}
\left({m_{\tilde k} \over 100 \ {\rm GeV}}\right)^2, 
\ee
\be
|(\lambda_{\cdots} \ {\rm or} \ \lambda'_{\cdots})
\lambda^{\prime\prime}_{\cdots}| < 10^{-10} \left({m_{\tilde k} \over
100 \ {\rm GeV}}\right)^2, 
\ee
\es
where $m_{\tilde k}$ is the mass of a squark of flavor $k$ (generic in
the case of 6c) and $\cdots$ stand for any any set of three flavor
indices.  Because (6) is so stringent, we will henceforth consider either
$L$-violating and $B$-conserving $\lambda,\lambda'$-couplings {\it or}
$B$-violating, $L$-conserving $\lambda^{\prime\prime}$-ones.
\bigskip

\noindent $\bullet$ {\bf Bounds on \R \ couplings}
\medskip

\nobreak
In deriving bounds on the absolute values of couplings, we will assume that
one \R \ operator acts at a time.  Known upper
bounds [9-14] on various $\lambda$- and $\lambda^{\prime\prime}$ type of
couplings, along with the physical process which is the source of each
bound, are listed in Table 1.  Specific references to the original
derivation of these bounds may be found in the review by Bhattacharyya
[10].  A similar listing has also been done, for whatever bounds [9-15] that 
exist on the $\lambda'$ couplings.  One remark on the bounds from
$D\bar D$ mixing and $D$-decays is in order.  In SM no separate
information on the unitary matrices $U^u_L$, $\bar U^d_L$,
transforming $u,d$ quarks to the mass-diagonal basis, is ever needed.
Only the product combination $U^{u^\dagger}_L U^d_L \equiv V_{CKM}$ is
constrained by experiment.  Such is not the case once \R \ interactions
are included.  The bounds above, pertaining to the $D$-system,
have been derived by assuming $U^d_L = 1$.
\newpage
\begin{center}
TABLE 1
\end{center}
\smallskip
\begin{enumerate}
\item[{}] Upper bounds on \R \ couplings for $\tilde m = 100$ GeV.
The numbers with $(\star)$ correspond to $2\sigma$ limits and those
with ($\ddagger$) are basis-dependent limits.
\end{enumerate}
\begin{center}
PART A
\end{center}
\[
\begin{tabular}{c l l c l l}
\hline
&&&&& \\
Coupling & Bound & Source & Coupling & 
Bound & Source \\
&&&&& \\
\hline
&&&&& \\
$|\lambda_{121}|$ & 0.05$^\star$ & CC univ. &
$|\lambda^{\prime\prime}_{112}|$ & $10^{-6}$ & $N N \rightarrow K$'s \\
$|\lambda_{122}|$ & 0.05$^\star$ & CC univ. &
$|\lambda^{\prime\prime}_{113}|$ & $10^{-5}$ & $n\bar n$ oscillation \\
$|\lambda_{123}|$ & 0.05$^\star$ & CC univ. &
$|\lambda^{\prime\prime}_{123}|$ & 1.25 & Pert. unitarity \\
$|\lambda_{131}|$ & 0.06 & $\Gamma(\tau \rightarrow e\nu\bar\nu)/\Gamma(\tau
\rightarrow \mu\nu\bar\nu)$ & $|\lambda^{\prime\prime}_{212}|$ & 1.25
& Pert. unitarity \\ 
$|\lambda_{132}|$ & 0.06 & $\Gamma(\tau \rightarrow e\nu\bar\nu)/\Gamma(\tau
\rightarrow \mu\nu\bar\nu)$ & $|\lambda^{\prime\prime}_{213}|$ & 125 &
Pert. unitarity \\ 
$|\lambda_{133}|$ & 0.003 & $\nu_e$-mass &
$|\lambda^{\prime\prime}_{223}|$ & 1.25 & Pert. unitarity \\ 
$|\lambda_{231}|$ & 0.06 & $\Gamma (\tau \rightarrow e\nu\bar\nu)/\Gamma(\tau
\rightarrow \mu\nu\bar\nu)$ & $|\lambda^{\prime\prime}_{312}|$ &
$10^{-5}$ & $n\bar n$ oscillation \\
$|\lambda_{232}|$ & 0.06 & $\Gamma(\tau \rightarrow e\nu\bar\nu)/\Gamma(\tau
\rightarrow \mu\nu\bar\nu)$ & $|\lambda^{\prime\prime}_{313}|$ &
$10^{-5}$ & $n\bar n$ oscillation \\
$|\lambda_{233}|$ & 0.06 & $\Gamma(\tau \rightarrow e\nu\bar\nu)/\Gamma(\tau
\rightarrow \mu\nu\bar\nu)$ & $|\lambda^{\prime\prime}_{323}|$ & 0.50
& $\Gamma^Z_\ell/\Gamma^Z_h$ (LEP 1) \\
\end{tabular}
\]
\begin{center}
PART B
\end{center}
\[
\begin{tabular}{c l l c l l c l l}
\hline
&&&&&&&& \\
Coupling & Bound & Source & Coupling & Bound & Source & Coupling &
Bound & Source \\
&&&&&&&& \\
\hline
$|\lambda'_{111}|$ & 0.00035 & $(\beta\beta)_{0\nu}$ &
$|\lambda'_{211}|$ & 0.09 & $R_\pi$ 
& $|\lambda'_{311}|$ & 0.10 & $\tau^- \rightarrow \pi^-
\nu_\tau$ \\ 
$|\lambda'_{112}|$ & 0.02$^\star$ & CC univ. & $|\lambda'_{212}|$ &
0.09 & $R_\pi$ 
& $|\lambda'_{312}|$ & 0.10 & $\tau^- \rightarrow \pi^-
\nu_\tau$ \\ 
$|\lambda'_{113}|$ & 0.02$^\star$ & CC univ. & $|\lambda'_{213}|$ &
0.09 & $R_\pi$ 
& $|\lambda'_{313}|$ & 0.10 & $\tau^- \rightarrow \pi^-
\nu_\tau$ \\ 
$|\lambda'_{121}|$ & 0.035$^\star$ & APV & $|\lambda'_{221}|$ & 0.18
& $D$-decay & 
$|\lambda'_{321}|$ & 0.20$^\ddagger$ & $D^0\bar D^0$ mix. \\
$|\lambda'_{122}|$ & 0.02 & $\nu_e$-mass & $|\lambda'_{222}|$ & 0.18 &
$D$-decay & 
$|\lambda'_{322}|$ & 0.20$^\ddagger$ & $D^0\bar D^0$ mix. \\
$|\lambda'_{123}|$ & 0.20$^\ddagger$ & $D^0 \leftrightarrow \bar D^0$ &
$|\lambda'_{223}|$ & 
0.18 & $D$-decay & $|\lambda'_{323}|$ & 0.20$^\ddagger$ & $D^0\bar
D^0$ mix. \\ 
$|\lambda'_{131}|$ & 0.035$^\star$ & APV & $|\lambda'_{231}|$ &
0.22$^\star$ & 
$\nu_\mu$ d.i. & $|\lambda'_{331}|$ & 0.48 & $R_\tau$ (LEP) \\
$|\lambda'_{132}|$ & 0.34 & $R_e$ (LEP) & $|\lambda'_{232}|$ & 0.36 &
$R_\mu$ & $|\lambda'_{332}|$ & 0.48 & $R_\tau$ (LEP) \\
$|\lambda'_{133}|$ & 0.0007 & $\nu_e$-mass & $|\lambda'_{233}|$ & 0.36
& $R_\mu$ & $|\lambda'_{333}|$ & 0.48 & $R_\tau$ (LEP) \\
&&&&&&&& \\
\hline
\end{tabular}
\]
\bigskip

An interesting recent development has been the derivation [16] of bounds on
products of different \R \ couplings that are more stringent than the
products of the individual bounds available.  These follow if, instead of
assuming the action of only one operator, we allow the possibility of
two different operators acting at two different vertices in a Feynman
diagram. Some of these bounds are tabulated below with the
corresponding processes cited.  Additional results may be found in the 
recent papers of Jang et al [16].

\newpage

\begin{center}
TABLE 2
\end{center}
\smallskip
\begin{enumerate}
\item[{}] Upper bounds on some important product couplings for $\tilde
m = 100$ GeV.
\end{enumerate}
\bigskip
\[
\begin{tabular}{l l l l l l}
\hline
&&&&& \\
Combination & Bound & Source & Combination & Bound & Source \\
&&&&& \\
\hline
&&&&& \\
$|\lambda'_{11k} \lambda^{\prime\prime}_{11k}|$ & $10^{-27}$ & Proton
decay & $|\lambda'_{ijk} \lambda^{\prime\prime}_{lmn}|$ & $10^{-10}$ &
Proton decay \\
&&&&& \\
$|\lambda_{1j1} \lambda_{1j2}|$ & $7.10^{-7}$ & $\mu \rightarrow 3e$ &
$|\lambda_{231} \lambda_{131}|$ & $7.10^{-7}$ & $\mu \rightarrow 3e$
\\
&&&&& \\
$|{\rm Im} \ \lambda'_{i12} \lambda^{\prime \star}_{i21}|$ &
$8.10^{-12}$ & $\epsilon_K$ & $|\lambda'_{i12} \lambda'_{i21}|$ &
$1.10^{-9}$ & $\Delta m_K$ \\
&&&&& \\
$|\lambda'_{i13} \lambda'_{i31}|$ & $8.10^{-8}$ & $\Delta m_B$ &
$|\lambda'_{1k1} \lambda'_{2k2}|$ & $8.10^{-7}$ & $K_L \rightarrow \mu
e$ \\
&&&&& \\
$|\lambda'_{1k1} \lambda'_{2k1}|$ & $5.10^{-8}$ & $\mu {\rm Ti}
\rightarrow e {\rm Ti}$ & $|\lambda'_{11j} \lambda'_{21j}|$ &
$5.10^{-8}$ & $\mu {\rm Ti} \rightarrow e {\rm Ti}$ \\
&&&&& \\
$|\lambda^{\prime\prime}_{i32} \lambda^{\prime\prime}_{i21}|$ & 0.008
& $\Gamma(B^{\rm ch} \rightarrow K^{neut.} K^{ch})$ &
$|\lambda^{\prime\prime}_{i31} \lambda^{\prime\prime}_{i21}|$ & 0.006
& $\Gamma(B^{ch} \rightarrow K^{neut.} \pi^{ch})$ \\
&&&&& \\
\hline
\end{tabular}
\]
\bigskip

\noindent $\bullet$ {\bf Collider-specific phenomenology}
\medskip

\nobreak
Signatures for \R \ interactions in collider experiments involve two
aspects of the concerned processes: first, production of some
sparticles and second their specific decay modes.  If sparticle
pair-production is utilized, as is usually the case, only MSSM
vertices are involved in the production process, whereas \R \ shows up
through the sparticle decays 
that follow.  Single sparticle production is, however, possible with
\R \ vertices and can be exploited sometimes.
\bigskip

\noindent {\it Tevatron $\rightarrow$ LHC}
\medskip

\nobreak
In such hadron colliders the possible production and decay-chains are
\[
q\bar q \ {\rm or} \ gg \rightarrow \tilde q \tilde q^\star
\rightarrow qq \tilde\chi^0_1 \tilde\chi^0_1,
\]
\[
\tilde\chi^0_1 \rightarrow \ell\bar\ell \nu \ {\rm or} \ q' \bar q\ell.
\]
Whether $\tilde\chi^0_1$ decays largely into $\ell^+ \ell^- \nu$ or
two jets and a lepton depends on which of the operators $L_i \cdot L_j
\bar E_k$ and $L_i \cdot Q_j \bar D_k$ in (2b) is dominant.  These
$\ell \ell \bar\ell \bar\ell$ \E \ or $jj jj \ell\ell$ signals
consistute very characteristic signatures [17] of \R.  Another process
involving \R \ interactions in both production and decay is
\[
g + d \rightarrow \tilde u_j + \ell^-,
\]
\[
\tilde u_j \rightarrow d_k + \bar\ell,
\]
leading to a $\ell\bar\ell j$ final state.  The lower limits of
$O(250) \ {\rm GeV}$, derived from Tevatron data on squark and gluino masses
in $MSSM$, get diluted to about $O(100-130) \ {\rm GeV}$ in \R \ $MSSM$.
Another important point is that $\lambda'$-couplings make additional
contributions to top semileptonic decay $t \rightarrow b \ell^+ \nu$,
leading to a violation of $e-\mu$ universality.
\bigskip

\noindent {\it HERA}
\medskip

\nobreak
Resonant squark production in deep inelastic $ep$ scattering is
possible via \R \ interactions, e.g.
\beq
e^+d &\rightarrow& \tilde u \rightarrow (e^+ d_j,\bar\nu u_j,\tilde
\chi^0_1,u_j, \tilde\chi_1^+ d_j), \nonumber \\[2mm]
e^- u &\rightarrow& \tilde d \rightarrow (\bar e u_j, \bar\nu_e
d_j, \tilde \chi^0_1 d_j, \tilde \chi^-_1 u_j), \nonumber \\[2mm] 
\tilde\chi^0_1 &\rightarrow& (\ell^\pm \ {\rm or} \ \nu) jj \ {\rm or}
\ jj\ell, \nonumber \\[2mm] \tilde\chi^\pm_1 &\rightarrow& (\ell^\pm \
{\rm or} \ \nu) jj \ {\rm or} \ jj\ell, \nonumber
\eeq
leading to a variety of final states containing multileptons + \E \ or
multileptons + jets.  For these signals to be observable, assuming
that the squark masses 
are within the energy reach, one will need the relevant $\lambda'$
couplings to be between [18] $O(10^{-2})$ and $O(10^{-1})$ in
magnitude.  A point to note is that these are precisely the
couplings that are relevant to atomic parity violation (cf. Table 1). 
\bigskip

\noindent {\it LEP 2 and NLC}
\medskip

\nobreak
In $e^+e^-$ colliders one can look for LSP pair-production followed by
\R \ decays:
\beq
e^+e^- &\rightarrow& \tilde\chi^0_1 \tilde\chi^0_1 \nonumber \\[2mm]
\tilde\chi^0_1 &\rightarrow& \ell^+ \ell^- \nu \ {\rm or} \
jj\ell. \nonumber 
\eeq
Thus one can again have quadrilepton + \E \ or dilepton + multijet final
states.  Particularly interesting [14] are final states containing
an isolated likesign ditau with multijets.  These signals can be
visible even if the concerned $\lambda$ or $\lambda'$ couplings are as
low as $O(10^{-4})$. 
\bigskip

\noindent $\bullet$ {\bf Cosmological implications}
\medskip

\noindent {\it Baryogenesis}
\medskip

\nobreak
Any baryon asymmetry, generated at temperatures pertaining to the GUT
scale, is in danger of being washed out by sphaleron-induced processes
if there are additional $B\!\!\!\!/ - L\!\!\!/$ interactions at the
electroweak scale.  The requirement of the preservation of primordial
baryon asymmetry was alleged to impose strong $(< 10^{-7})$
constraints on the strengths of \R \ interactions which violate $B\!\!\!\!/
- L\!\!\!/$.  However, the conservation of ${1\over3} B - L_i$, where
$i$ is any $B\!\!\!\!/ - L\!\!\!/$ {\it one} generation, has been shown [19]
to suffice for this purpose -- thereby leaving most of the \R \ couplings
cosmologically unconstrained.
\newpage

\noindent {\it LSP decay}
\medskip

\nobreak
The simplest possible decay mode of the lightest neutralino is
$\tilde\chi^0_1 \rightarrow \bar e u \bar d$ with the rate given by
Dawson's formula [20] $\tau^{-1}_{\tilde\chi_0} = (128 \pi^2)^{-1}
3\alpha |\lambda'_{121}|^2 M^5_{\tilde\chi^0_1} m^{-4}_{\tilde e}$.
The requirement of the LSP decaying within a detector can be
quantified as $c\gamma_L \tau_{\tilde\chi_0} < 1 m$ and leads to the
constraint 
\[
|\lambda'_{121}| > 1.4 \times 10^{-6} \sqrt{\gamma_L} \left({m_{\tilde
e} \over 200 \ {\rm GeV}}\right)^2 \left({100 \ {\rm GeV} \over
M_{\tilde\chi}}\right)^{5/2}, 
\]
where $\gamma_L$ is the Lorentz boost factor.  On the other hand, from
the experimental absence of a long-lived relic LSP whose decay would
lead to detectable upward going muons, one can infer [21] that $1 \
{\rm s} <
\tau_{\tilde\chi} < 10^{17}$ yrs leading to the forbidden interval
$10^{-10} \not< |\lambda,\lambda',\lambda^{\prime\prime}| \not<
10^{-22}$.  Note that $\tau_{\tilde\chi} > 10^{17}$ yrs would make \R
\ $MSSM$ indistinguishable from $MSSM$.  Furthermore, the interval
$10^{-8} s < \tau_{\tilde\chi_1} < 1s$ is practically out of
observational reach.  Of course, with an unstable LSP, one needs
another candidate particle (axion?) for cold dark matter. 
\bigskip

\noindent {\bf Summary and outlook}
\medskip

\nobreak
We conclude by emphasizing four main points.  (1) There is presently
no credible theoretical objection against the presence of \R \
interactions near the weak scale.  (2) Though many of the \R \ couplings
have been constrained in Tables 1-3, several (e.g. $\lambda'_{322},
\lambda'_{323}$ etc.) are totally unconstrained.  (3) Vigorous seaches
are in progress and detection strategies are under formulation 
for collider signatures of \R \ in terms of excess
lepton-jet combinations at HERA, violation of $e - \mu$ universality
in top semileptonic decay at the Tevatron and isolated ditaus and jets
in LEP 2 and NLC.  (4) It is highly desirable to find some way of
discriminating between different CDM candidates such as a neutralino
and an axion.

I compliment the P$^4$ workshop organizers for a job well-done.
\bigskip

\noindent $\bullet$ {\bf References}
\smallskip

\begin{enumerate}
\item[{[1]}] H.E. Haber and G.L. Kane, Phys. Rep. {\bf 117} (1975) 75.
\item[{[2]}] G. Farrar and P. Fayet, Phys. Lett. {\bf B76} (1978) 575. 
\item[{[3]}] L.E. Ibanez and G.G. Ross, Phys. Lett. {\bf B260} (1991);
Nucl. Phys. {\bf B368} (1992) 3.  T. Banks and M. Dine,
Phys. Rev. {\bf D45} (1992) 1424.
\item[{[4]}] L.J. Hall and M. Suzuki, Nucl. Phys. {\bf B231} (1984) 419.
\item[{[5]}] D. Brahm and L.J. Hall, Phys. Rev. {\bf D40} (1989)
2449.  K. Tamvakis, Phys. Lett. {\bf B383} (1996) 207.  G.F. Giudice
and R. Rattazzi, hep-ph/9704339.
\item[{[6]}] M.C. Bento, L.J. Hall and G.G. Ross, Nucl. Phys. {\bf
B292} (1987) 400.   
\item[{[7]}] C.S. Aulakh and R.N. Mohapatra, Phys. Lett. {\bf 119B}
(1982) 136.  A. Santamaria and J.W.F. Valle, Phys. Lett. {\bf 195B}
(1987) 423.
\item[{[8]}] R. Hempfling, Nucl. Phys. {\bf B478} (1996) 3.
B. Mukhopadhay and S. Roy, Phys. Rev. {\bf D55} (1997) 7020. 
\item[{[9]}] J.L. Goity and M. Sher, Phys. Lett. {\bf B346} (1995)
69.  A.Yu. Smirnov and F. Vissani, Phys. Lett. {\bf B380} (1996) 317.  
\item[{[10]}] V. Barger, G.F. Giudice and T. Han, Phys. Rev. {\bf D40}
(1989) 2987.  G. Bhattacharyya, hep-ph/9709396.  H. Dreiner,
hep-ph/9707435.   
\item[{[11]}] B. Brahmachari and P. Roy, Phys. Rev. {\bf D50} (1994)
39, err. {\it ibid.} {\bf D51} (1995) 193. 
\item[{[12]}] D. Chang and W-K. Keung, Phys. Lett. {\bf B389} (1996) 294.
\item[{[13]}] G. Bhattacharyya, D. Choudhury and K. Sridhar,
Phys. Lett. {\bf B355} (1995) 193.
\item[{[14]}] R.M. Godbole, P. Roy and X. Tata, Nucl. Phys. {\bf B401}
(1993)67.
\item[{[15]}] M. Hirsch, H.V. Klapdor-Kleingrothaus and
S.G. Kovalenko, Phys. Rev. Lett. {\bf 75} (1995) 2276.  K.S. Babu and
R.N. Mohapatra, Phys. Rev. Lett. {\bf 75} (1995) 2276. 
\item[{[16]}] D. Choudhury and P. Roy, Phys. Lett. {B378} (1996) 153.
J-H. Jang, J.K. Kim and J.S. Lee, Phys. Rev. {\bf D55} (1997) 7296.
\item[{[17]}] H. Dreiner and G.G. Ross, Nucl. Phys. {\bf B365} (1991)
597.  D.P. Roy, Phys. Lett. {\bf B128} (1992) 270.
\item[{[18]}] D. Choudhury and S. Raychaudhuri, Phys. Lett. {\bf B401}
(1997) 54.
\item[{[19]}] H. Dreiner and G.G. Ross, Nucl. Phys. {\bf B410} (1993) 183.
\item[{[20]}] S. Dawson, Nucl. Phys. {\bf B261} (1985) 297.
\item[{[21]}] E.A. Baltz and P. Gondolo, hep-ph/9704411. 
\end{enumerate}

\end{document}